\documentclass{article}
\usepackage{emulateapj,apjfonts}
\usepackage{lscape}
\input psfig.sty

\lefthead{NGC 4565 AND NGC 5746: STRUCTURAL ANALOGS OF OUR GALAXY}
\righthead{KORMENDY AND BENDER}

\begin{document}

\centerline{\null}\vskip -40pt

\newcommand{\etal}{et~al.\ }
\newcommand{\ts}{\thinspace}

\def\gapprox{$_>\atop{^\sim}$} 
\def\lapprox{$_<\atop{^\sim}$}

\newdimen\sa  \def\sd{\sa=.03em  \ifmmode $\rlap{.}$''$\kern -\sa$
                                \else \rlap{.}$''$\kern -\sa\fi}

\newdimen\sb  \def\md{\sb=.04em \ifmmode $\rlap{.}$'$\kern -\sb$
                                \else \rlap{.}$'$\kern -\sb\fi}

\title{Structural Analogs of the Milky Way Galaxy:\\Stellar Populations in 
       the Boxy Bulges of NGC 4565 and NGC 5746\altaffilmark{1}}   
\author{John Kormendy\altaffilmark{2,3,4} and Ralf Bender\altaffilmark{3,4,2}}   

\altaffiltext{1}{Based on observations obtained with the Hobby-Eberly Telescope, 
                 which is a joint project of the University of Texas at Austin, 
                 the Pennsylvania State University, 
                 Ludwig-Maximilians-Universit\"at M\"unchen, and 
                 Georg-August-Universit\"at G\"ottingen.} 

\altaffiltext{2}{Department of Astronomy, University of Texas at Austin,
                 2515 Speedway, Mail Stop C1400, Austin,
                 Texas 78712-1205; kormendy@astro.as.utexas.edu}

\altaffiltext{3}{Universit\"ats-Sternwarte, Ludwig-Maximilians-Universit\"at, 
                 Scheinerstrasse 1,
                 M\"unchen D-81679, Germany}

\altaffiltext{4}{Max-Planck-Institut f\"ur Extraterrestrische Physik,
                 Giessenbachstrasse, D-85748 Garching-bei-M\"unchen, Germany; 
                 bender@mpe.mpg.de}

\begin{abstract} 
\noindent We present NGC 4565 and NGC 5746 as structural analogs of our Milky Way.
All three are giant, SBb\ts--{\ts}SBbc galaxies with two pseudobulges, 
i.{\ts}e., a compact, disky, star-forming pseudobulge embedded in a vertically
thick, ``red and dead'', boxy pseudobulge that really is a bar seen almost end-on. 
The stars in the boxy bulge of our Milky Way are old and enhanced in $\alpha$ 
elements, indicating that star formation finished within $\sim$\ts1 Gyr of when
it started.  Here, we present Hobby-Eberly Telescope spectroscopy of the boxy 
pseudobulges of NGC 4565 and NGC 5746 and show that they also are made of old 
and $\alpha$-element-enhanced stars.  Evidently it is not rare that the 
formation of stars that now live in bars finished quickly and early, even in 
galaxies of intermediate Hubble types whose disks still form stars now. Comparison 
of structural component parameters leads us to suggest that NGC\ts4565 and NGC\ts5746
are suitable analogs of the Milky Way, because they show signatures of similar
evolution processes.
\end{abstract}


\section{Introduction}

\pretolerance=15000  \tolerance=15000

      Physical analogs of unique objects such as our Sun~or~our 
Galaxy{\ts}are{\ts}important{\ts}because{\ts}they{\ts}illuminate{\ts}evolution{\ts}histories. Nature thus
presents us with an important opportunity, because the edge-on spiral 
galaxies NGC\ts4565 and NGC\ts5746 (Fig.\ts1) are close structural analogs 
of our Milky Way.  This conclusion is developed here in Section 4.

      All three galaxies are giants with outer, circular-orbit rotation velocities
$\sim$\ts244 km s$^{-1}$ (NGC\ts4565:~Hyperleda), 
$\sim$\ts311 km~s$^{-1}$ (NGC\ts5746:~Hyperleda), and 
$\sim$\ts220\ts$\pm$\ts20 km s$^{-1}$ (Milky Way: Bland-Hawthorn \& Gerhard 2016).~All 
are disk-dominated galaxies with ``boxy bulges''.  
It is well known that these are not classical, elliptical-galaxy-like
bulges.  Rather, they are bars seen nearly 
end-on.~Maihara{\ts}et{\ts}al.\ts(1978), 
Weiland{\ts}et{\ts}al.\ts(1994), and
Dwek \etal (1995) present the observations of our Galaxy, and
Blitz \& Spergel (1991) emphasize the robust argument that its 
parallelogram-shaped structure is a perspective effect caused by
the fact that the near side of the bar is closer~to~us~than~the~far 
side.~N-body simulations show that bars buckle vertically into boxy structures
(Combes \& Sanders 1981;
Combes \etal 1990;
Pfenniger \& Norman 1990;
Pfenniger \& Friedli 1991;
Raha~et al.\ts1991;
Athanassoula \& Misiriotis 2002;
Athanassoula 2005; 
Shen \etal 2010).
Observed boxy bulges show kinematic signatures of 
edge-on bars\ts--{\ts}cylindrical rotation 
(NGC 4565: Kormendy \& Illingworth 1982; 
our Galaxy:~Howard \etal 2008, 2009; 
see Kormendy \& Barentine 2010 for further review) and
a figure-8-shaped splitting of spectral emission lines indicative of gas 
flow in edge-on bars 
(NGC 5746:~Kuijken \& Merrifield 1995;
           Merrifield 1996;
           Merrifield \& Kuijken 1999; 
           Bureau \& Freeman 1999). 
If NGC 4565, NGC 5746, or our Galaxy were viewed face-on, then their 
``boxy bulges'' would be called bars.  None of these components would 
be confused with a bulge.

      Do these galaxies also contain classical bulges?  The answer is ``no'' 
in our Galaxy and ``not significantly'' in NGC\ts4565 and NGC\ts5746 
(\S\ts4).  From a hierarchical clustering point of view, these
are pure-disk galaxies with dynamically quiet  histories dominated by minor mergers
and bar-driven secular evolution for most of the history of the Universe 
(Kormendy \etal 2010).

      ``Secular evolution'' is so named because it is slow -- it has 
characteristic time scales of many galactic rotations, as distinct from 
fast processes that occur on galaxy~collapse~time~scales.  Still,
ample evidence for pseudobulges in ``red and dead'' S0 galaxies indicates 
that it can operate effectively for a few billion years in the early 
Universe, when gas was plentiful, and still~be star-formation-quenched 
long ago (Kormendy \& Kennicutt 2004; Kormendy 2013).  It is nevertheless 
at least superficially counter-intuitive that the boxy pseudobulge of our 
Galaxy is believed to be made almost exclusively of old and
$\alpha$-element-enhanced stars
(Renzini 1994;
Ortolani \etal 1995;
Matteucci \& Romano 1999;
Kuijken \& Rich 2002;
Zoccali \etal 2003; 
Clarkson \etal 2008; 
Ness \etal 2013; see
Rich 2013 and
Barbuy \etal 2018 for reviews, and
Renzini 2006;
Conroy \etal 2014, and
Nataf 2016
for cautions).  
The observation of $\alpha$ element enhancement particularly points to a 
short, $<$\ts1{\ts}Gyr star formation time scale. 
This may be reconciled with the Galaxy's barred structure if the stars 
formed first, long ago, and the bar grew secularly only
later (Freeman~2008).

      Is our Galaxy's old, $\alpha$-element-enhanced boxy bulge unique?  The purpose 
of this paper is to measure the stellar populations
in the well known, prototypical boxy bulges of NGC 4565 and NGC 5746 as close
analogs of our Galaxy.  We find that they, too, are made of old stars that 
are enhanced in $\alpha$ elements.  

\vskip -15pt
\centerline{\null}

\section{Hobby-Eberly Telescope Spectroscopy}

NGC 4565 and NGC 5746 were observed in February and May of 2004 with the 
Hobby-Eberly Telescope (Ramsey 
\etal 1998) and the Marcario Low Resolution Spectrograph (Hill \etal 1998) 
in service mode. % in February and May 2004. 
We used a 1\sd\kern 0.1pt0 wide
$\times$ 3\kern 0.5pt\md5 long slit, the G2 grism with 600 lines mm$^{-1}$, 
and a Ford Aerospace CCD detector binned 2$\times$2 yielding 
1568$\times$512 pixels and a spatial scale of 0{\kern 0.5pt}\sd{\kern 0.05pt}47 
pixel$^{-1}$.  The wavelength range covered is 4300\ts\AA~to~7200 \AA~with 
a median instrumental resolution~of $\sigma_{\rm instr} 
\approx 120$ km s$^{-1}$~Data reduction followed Saglia{\ts}et{\ts}al.\ts(2010).

Individual exposure times were 1800 s; total exposure times for
coadded images are given in the Figure 1 key. 
We also observed 16 standard stars from Worthey \etal
(1994) to enable calibration to the Lick system of spectral line indices
(Faber \etal 1985, Burstein~et~al.~1986).  Stellar types ranged from
G5 to K4 and included dwarf and giant stars.  Standard stars covered 
Mgb equivalent widths between 2\ts\AA~and 7\ts\AA~and <Fe> $\equiv$ 
(Fe $\lambda$5227 + Fe $\lambda$5335)/2 equivalent widths between 1.5\ts\AA~and 
4.5\ts\AA. Only minor corrections of $<$\ts5\ts\%~were needed
to calibrate our line measurements to the Lick system.

Kinematic parameters (velocities, velocity dispersions, and Gauss-Hermite
moments) were derived with the Fourier correlation quotient method
(Bender 1990; Bender{\ts\ts}et{\ts}al.\ts1994).

      Line indices were corrected for velocity  
broadening and brought to the standard resolution of the Lick system.

Figure 1 illustrates our spectrograph slit positions.  

\vskip 4.05truein

\includegraphics{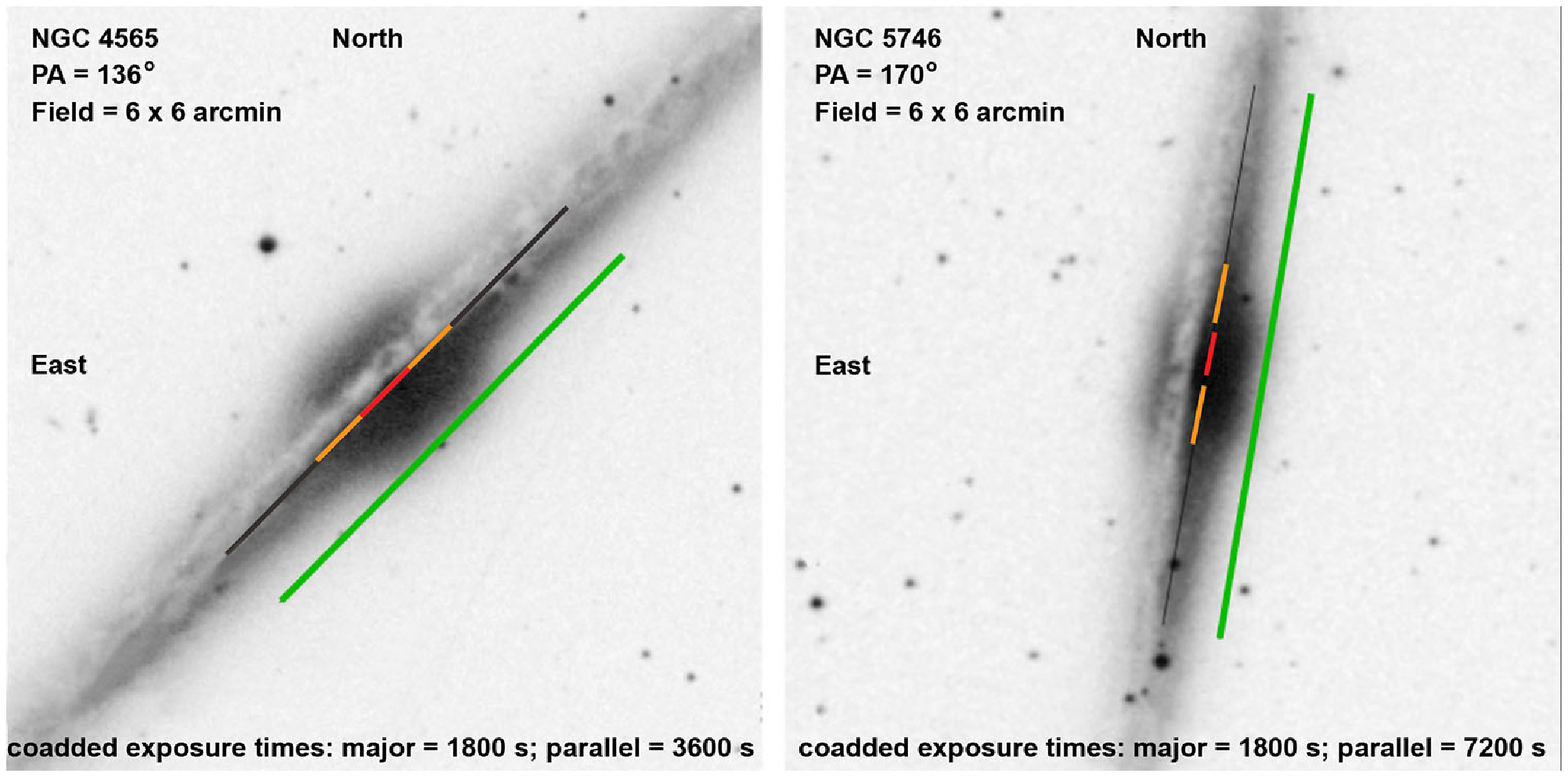}

\includegraphics{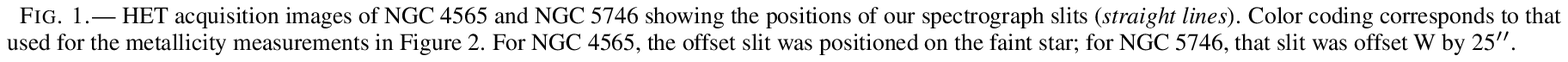}

\vskip 3.80truein

\includegraphics{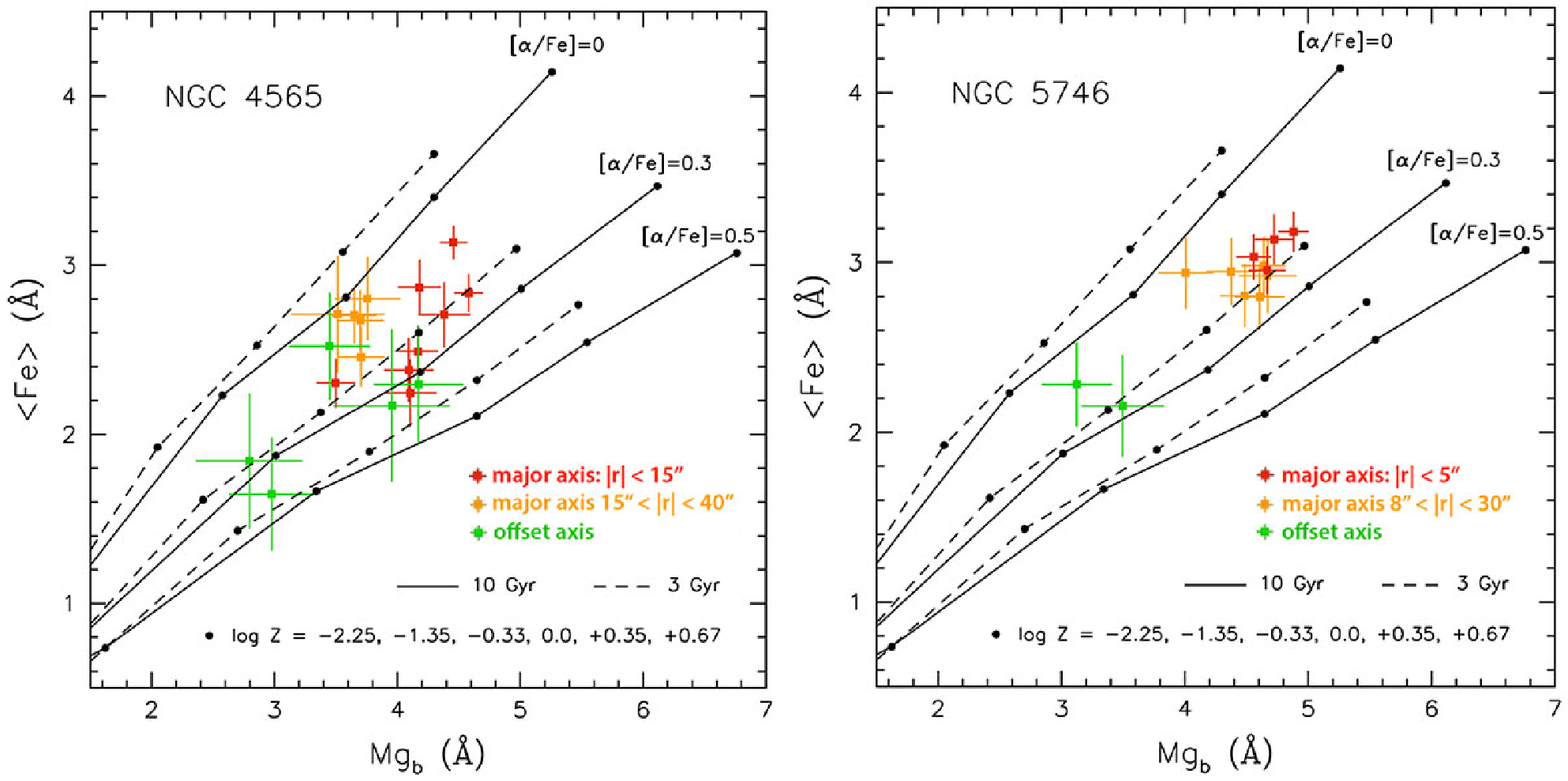} 

\includegraphics{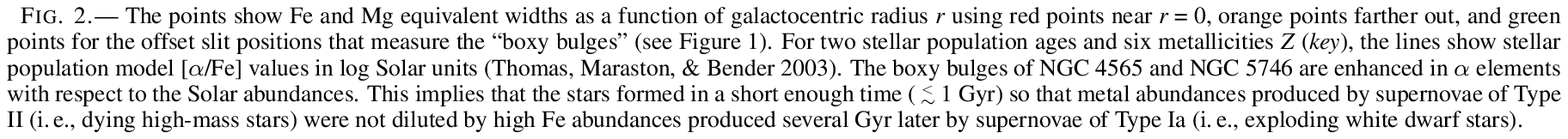}

\section{Element Abundance Patterns}

Figure 2 compares our results with the stellar population models of Thomas, 
Maraston, \& Bender (2003).  Both galaxies show metallicity gradients with
the highest metallicities near their centers. 
Stellar populations are not young at any radius, though intermediate-age 
and old populations cannot be distinguished.  Our important result is this:
Both pseudobulges -- and, indeed, even the inner disks of these galaxies -- 
are enhanced in $\alpha$ elements with respect to the Solar value by about 
a factor of two.  As in our Galaxy's almost-end-on, boxy bar, the star formation
in the bars of NGC\ts4565 and NGC\ts5746 took place over \lapprox \ts1 Gyr
(Thomas \etal 2005 and references therein).

\vfill

\newpage

\section{NGC 4565 and NGC 5746 as Milky Way Analogs}

      Sections 2 and 3 show that the almost-end-on, boxy bars of 
NGC\ts4565{\ts}and{\ts}NGC\ts5746{\ts}resemble{\ts}the{\ts}similarly{\ts}\hbox{almost-end-on}
bar in our Galaxy in stellar populations: all three are made of relatively old stars 
whose alpha elements are enhanced in abundances as compared to solar values.  
Evidently, in all three cases, the star formation that made the bars happened quickly 
and never got diluted later by star formation like that in the outer disks.  
From a stellar population point of view, NGC 4565 and NGC 5746 are good analogs 
of our Galaxy.  Close similarity is important in identifying analogs; the intent
is that study of the analogs illuminates our understanding of (in this case)
our own Galaxy.  This section emphasizes that NGC 4565 and NGC 5746 are also 
close structural analogs.

In fact, we suggest that they are more useful than the galaxies that are 
commonly chosen as Galaxy analogs, first because they \phantom{000000000000}

\vskip 6.95truein

\includegraphics{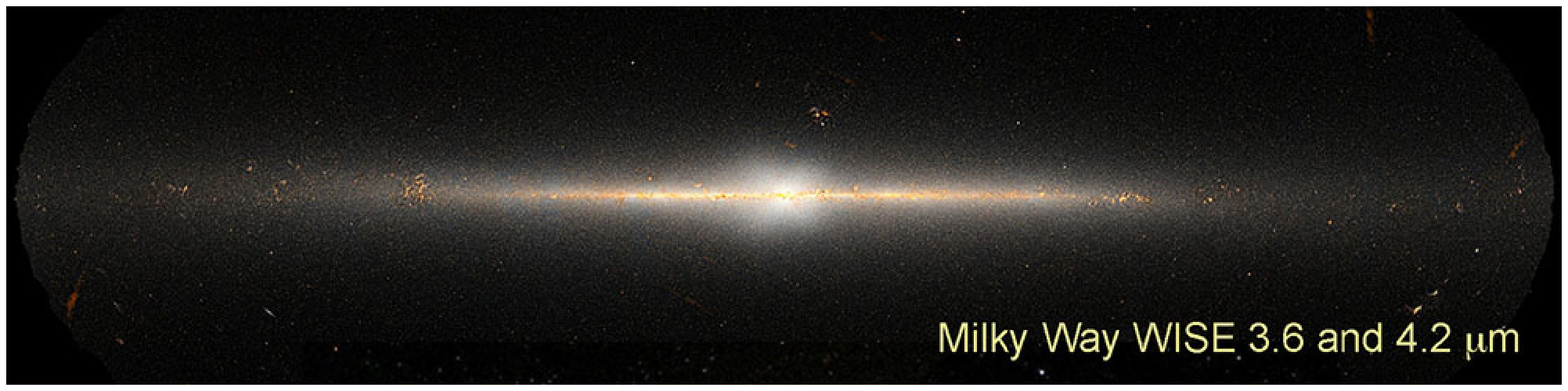}

\includegraphics{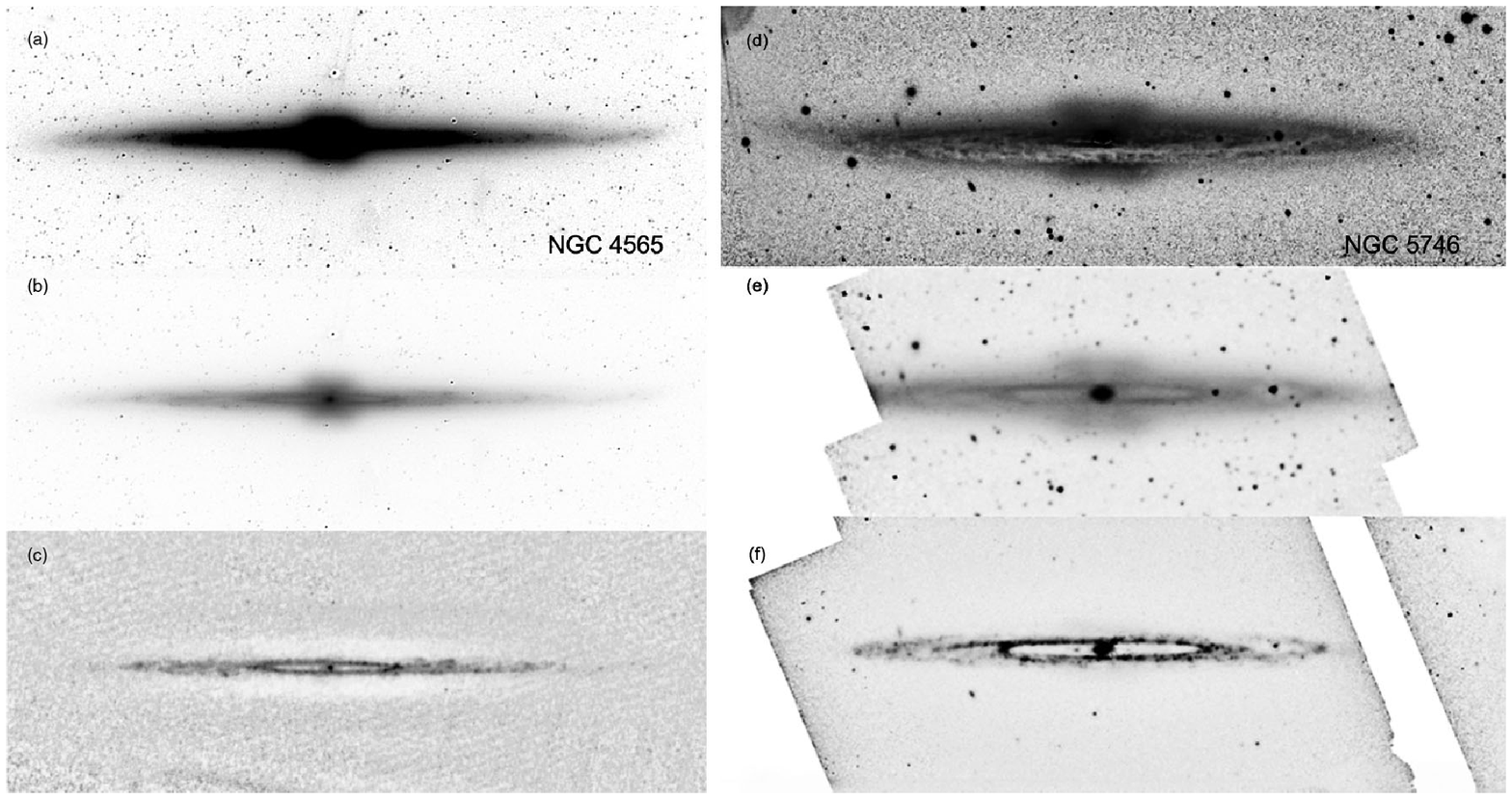}

\includegraphics{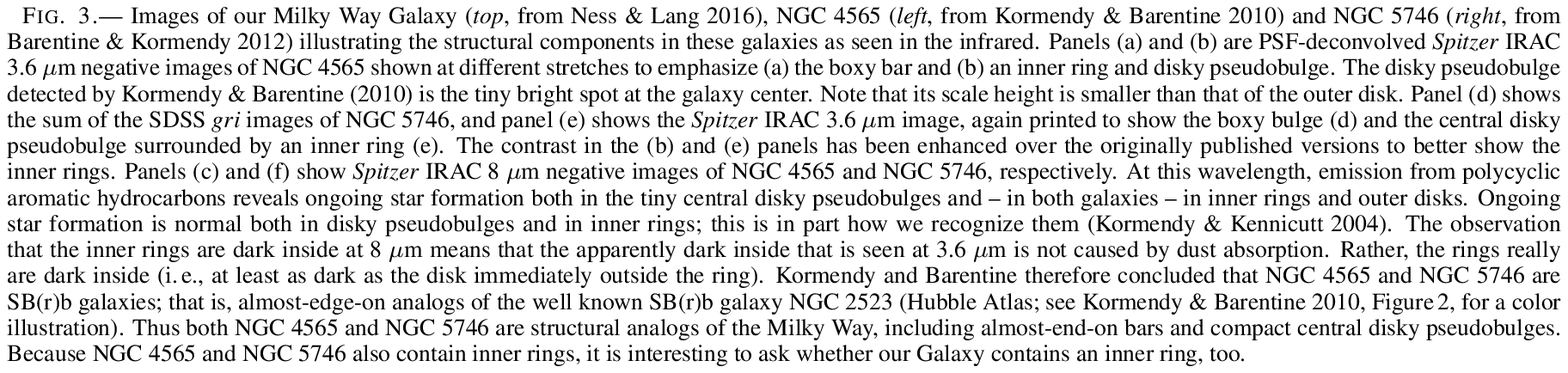}

\noindent focus attention on features that are the subjects of current study 
and second because they show those features cleanly and with high amplitude.
Galaxy analogs have generally been chosen to have similar spiral structure and 
either \hbox{bulge-to-total} mass ratios or bars.~Galaxies that have similar bulge 
masses but no bars are inappropriate:~our Galaxy contains no classical~bulge 
(see below).~Galaxies with similar bars are valid analogs, but our Galaxy may have a
somewhat weak bar, and perhaps in consequence, it has irregular spiral structure.
NGC 4565 and NGC 5746 are slightly more massive than our Galaxy and have similar
bar mass fractions.  Dynamics associated with bars, including all aspects 
of disk secular evolution (Kormendy \& Kennicutt 2004; Kormendy 2013) are enhanced. 
It is easier to study their cleaner dynamics.  On this basis, we suggest that 
these galaxies are practically the most useful Galaxy analogs.
Figure 3 and Table 1 make these points more concrete.

\eject

      Table 1 quantifies features shown in Figure 3 and measured in 
Kormendy \& Barentine (2010); Barentine \& Kormendy~(2012).  NGC 4565 and 
NGC 5746 are giant galaxies, slightly bigger and more massive than our Galaxy.
High mass is an advantage: structural components are bigger and the dynamics are 
cleaner and easier to measure with respect to measurement errors.

      Bland-Hawthorn \& Gerhard (2016) substantially advance the state of the art
in their review of the properties~of~our~Galaxy.  They find that the stellar mass 
of the ``bulge~region'' (we assume that this is the box structure) is 
(1.4{\ts}to{\ts}1.7) $\times$ $10^{10}$ $M_\odot$.  This is $0.3 \pm 0.06$
of the total stellar mass of our Galaxy.  Within 
measurement errors, this is the same as the luminosity ratios Boxy/T $\sim$ 0.3
measured for NGC 4565 and NGC 5746 (Table\ts1 notes). 
But~\hbox{$n$-body} models show that
vertically thick, boxy (when seen edge-on) parts of bars are only part of the
whole bar: they are embedded in vertically~thin~bars~that~are~radially~longer 
(Combes{\ts}et{\ts}al.\ts1990;{\ts}Raha{\ts}et{\ts}al.\ts1991;{\ts}Athanassoula\ts2005,\ts2013;
Shen \etal 2010).
Bland-Hawthorn \& Gerhard estimate that the Galaxy's thin bar has a stellar mass
of $(7 \pm 1) \times 10^9$ $M_\odot$.  Subtracting the nuclear disk mass, the 
complete bar then has
a bar-to-total stellar mass ratio of \hbox{Bar/T $\simeq$ $0.42 \pm 0.09$.}  
This is close to the largest Bar/T ratio measured in face-on barred galaxies. 
So the dynamical processes 
associated with bars are expected to be clean and strong in all three galaxies.

\vskip 5.9truein

 \includegraphics{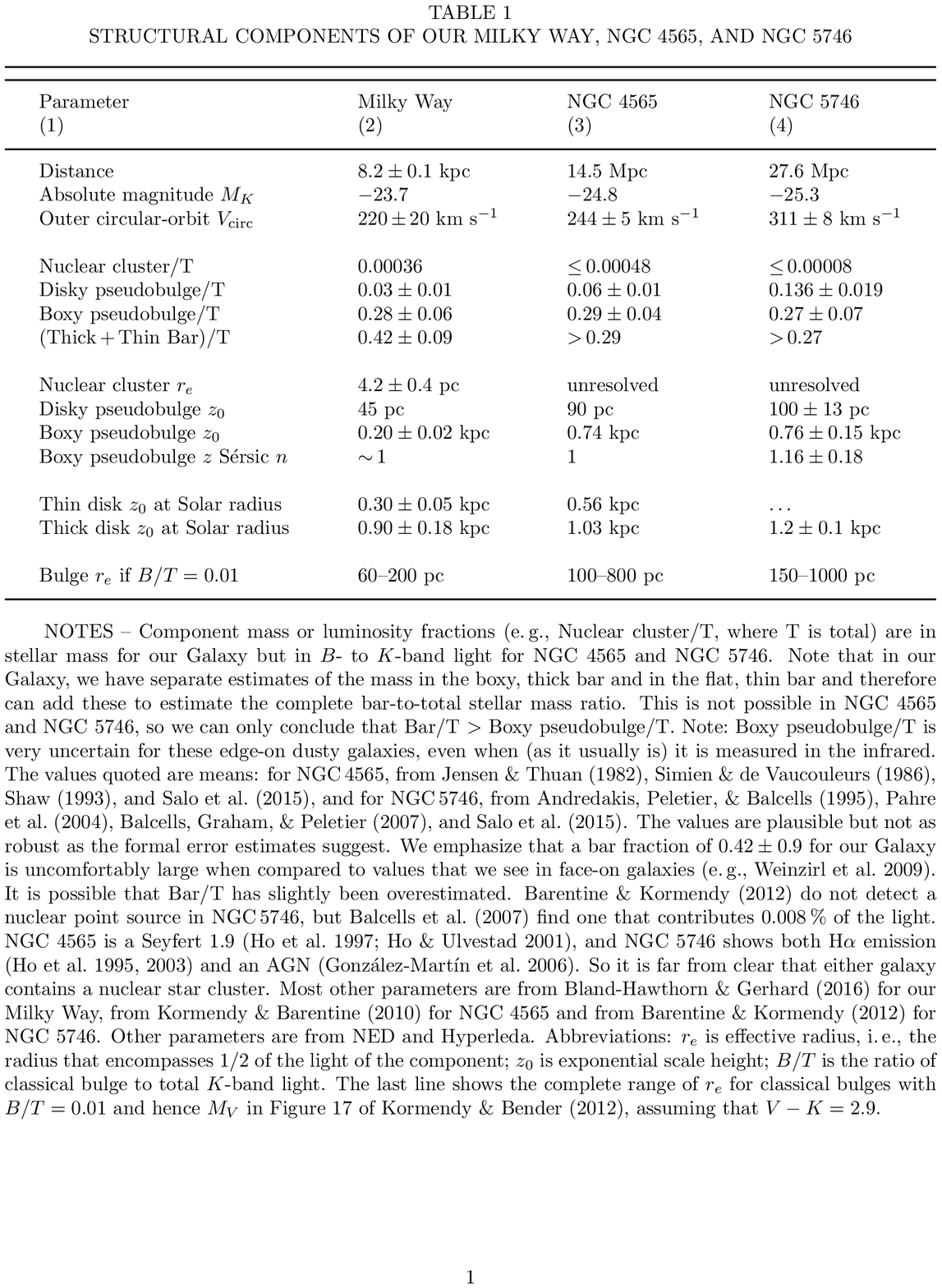}

      Important to our confidence that components are interpreted correctly
is the observation that the boxy bars (``pseudobulges'', 
because{\ts}they{\ts}form{\ts}from{\ts}disks) have exponential vertical density 
profiles (S\'ersic 1968 index $n \simeq 1$).  Note that the scale height of the boxy 
structure in our Galaxy is smaller than those in NGC 4565 and NGC 5746. 
Only in the external galaxies is this scale height larger than that of the 
thin~disk.  It is possible that the bar strength in our Galaxy is overestimated.  
Given uncertainties in bulge-disk decomposition and given internal absorption 
in edge-on galaxies, Bar/T is uncertain in all three objects.  
Nevertheless, the boxy structure is clearcut in all three.

     Relevance of NGC\ts4565 and NGC\ts5746 as Galaxy analogs depends on a
comparison of further evolution-related structural features.
An especially important aspect of bar-driven secular evolution is that it drives
some disk gas toward the center where it forms stars and builds
``disky pseudobulges'' (Kormendy \& Kennicutt 2004; Kormendy 2013 provide
reviews).  SB(r) galaxies such as NGC\ts4565 and NGC\ts5746 are relatively
mature in that most gas has been evacuated inside the inner ring (Fig.~3).
A sign of this is that inner ring galaxies almost never have dust lanes on 
the rotationally leading side of bars (Sandage 1975).  These are 
interpreted (Athanassoula 1992) as the signature of gas shocks
(e.{\ts}g., Regan \etal 1997) that result from non-circular forces produced
by bars and that result in gas \phantom{000000000000}

\vfill\eject

\noindent flow toward the center.  SB(r) galaxies
are dynamically more mature than SB(s) galaxies in which spiral arms begin
near the ends of bars (Sanders \& Tubbs 1980).  In them, pseudobulge
building is more nearly finished.  All three present galaxies are well known 
to be making stars, but only in small volumes near their centers (e.{\ts}g., 
Fig.~3).  These still-growing, disky pseudobulges have parameters listed 
in Table 1.  Our Galaxy has a small disky pseudobulge that is clearly distinct 
from the boxy pesudobulge.  It accounts for 
0.03 $\pm$ 0.01 of the total mass and is well known to show ongoing star 
formation (e.{\ts}g., Yusef-Zadeh \etal 2009).  Similar disky pseudobulges in 
NGC\ts4565 and NGC\ts5746 have 0.06 $\pm$ 0.01 and 0.14 $\pm$ 0.02 of the 
stellar masses.  All three have scale heights $z_0 \simeq 45$ to 100 pc that
are much smaller than the thin disk scale heights.  This -- not apparent 
flattening -- plus their ongoing star formation out of disk gas is why we call 
them ``disky pseudobulges''.

     Photometric decomposition of edge-on and face-on galaxies may differently
distinguish boxy and disky pseudobulges.~We 
may{\ts}overestimate{\ts}Bar/T{\ts}and{\ts}underestimate{\ts}(Disky{\ts}pseudobulge)/T
in all three galaxies,
as compared with how we would carry 
out photometric decomposition of face-on galaxies.

     Our Galaxy also contains a tiny nuclear star cluster:
(Nuclear cluster)/Total mass $\simeq$ 0.00036.  A central point source of light
in NGC 4565 contributes 0.00048 of the light but could be an AGN (Ho \etal 1997;
Ho \& Ulvestad 2001).  
NGC 5746 is almost twice as far away as NGC 4565; a nuclear cluster is not securely
detected.  Nuclear star clusters generally consist of a mixture of old and 
young stars and often show star formation.  Star formation in our
Galaxy's nucleus is discussed, e.{\ts}g., in Feldmeier-Krause \etal (2015)
and reviewed in Genzel \etal (2010).  Thus the growth of nuclear star clusters
must be secular.  Their origin is not well understood.  When there is a substantial
bulge, dynamical friction can deliver globular clusters to the galaxy center
(Tremaine \etal 1975). This is less easy in the lower-density Galactic bar. 
It would be interesting to check whether either galaxy contains a nuclear star cluster.

     Do these galaxies contain classical bulges?~The answer is ``no'' in our Galaxy 
and ``not significantly'' in NGC\ts4565 and NGC\ts5746.{\ts}Kormendy{\ts}et{\ts}al.\ts2010~\hbox{emphasize}:~{\it ``we~do~not~have~the freedom to postulate classical 
bulges which have arbitrary properties (such as low surface brightnesses)
that make them easy to hide.~Classical bulges and ellipticals satisfy 
well-defined fundamental plane correlations} (Kormendy et{\ts}al.\ts2009 and 
Kormendy\ts2009 show these to the faintest \hbox{luminosities}).''  The smallest 
dwarf true elliptical (VCC\ts1199: Kormendy \etal 2009) could be hidden in 
either NGC\ts4565 or NGC\ts5746; it would imply bulge-to-total $K$-band
luminosity~ratios $B/T \simeq 0.003$ and 0.002, respectively.  Classical bulges
with $B/T = 0.01$ would be hard to hide: their effective radii would 
clobber the tiniest structures seen in these galaxies (Table\ts1).   
Bigger bulges would be seen.  For our Galaxy, a stronger conclusion is possible.  
The center is so close that there is no room to hide even the tiniest known classical
bulge:~the high surface brightness and large effective radius predicted from 
fundamental plane correlations (Kormendy \etal 2009; Kormendy \& Bender 2012) would 
overwhelm the nuclear star cluster (Table 1).

      From a hierarchical clustering point of view, NGC\ts4565, NGC\ts5746, 
and our Galaxy are astrophysically similar, giant, and essentially pure-disk 
galaxies whose histories have been dominated by minor mergers and bar-driven 
secular evolution for most of the history of the Universe (Kormendy \etal 2010).

      A final question is motivated by Figure 3.  NGC~4565 and NGC 5746
clearly contain ``inner rings'' (de Vaucouleurs 1959; Sandage 1961; Buta \etal 2007)
that, in face-on galaxies, surround the ends of bars.  Given other similarities 
between these galaxies and our own, it is worth asking whether our Galaxy also 
could be an SB(r)bc galaxy.  Distances to gas concentrations within our Galaxy
are usually estimated based on measured line-of-sight velocities and the
assumption of {\it circular orbits.}  But a barred galaxy has significant 
noncircular rotation.  Could the ``3-kpc arm'' of our Galaxy be an inner ring, 
as suggested by Sevenster \& Kalnajs (2001)?  
Could some other feature be such a ring?  
No conclusions of this paper depend on the answer, but it would be an 
interesting addition to our understanding of the Milky Way.

\null

\acknowledgements 

      The spectra were taken with the Marcario Low-Resolution Spectrograph 
(LRS) and the Hobby-Eberly Telescope (HET).  LRS is named for Mike Marcario 
of High Lonesome Optics; he made the LRS optics but died before its completion.  
LRS is a project of the HET partnership and the Instituto de Astronom\' \i a 
de la Universidad Nacional Aut\'onoma de M\'exico.  The HET partners are the 
University of Texas at Austin, Pennsylvania State University, 
Ludwig-Maximilians-Universit\"at M\"unchen, and Georg-August-Universit\"at,
G\"ottingen.  The HET is named in honor of its principal benefactors,
William P.~Hobby and Robert E.~Eberly. 

      Much of this work was done while J.K.~was supported by the Curtis T.~Vaughan,
Jr.~Centennial Chair in Astronomy at the University of Texas at Austin. We are most 
sincerely grateful to Mr.~and Mrs.~Curtis T.~Vaughan, Jr.~for their many years of
support for Texas astronomy.  J.K.~was also supported by NSF grant AST-0607490.
R.B.~is grateful to the Department of Astronomy of the University of Texas 
at Austin for the Tinsley Visiting Professorship during which much of this work 
was done.  We thank the referee for a prompt and helpful report.  We also thank 
Mike Rich for helpful correspondence on the ages of Galactic bulge stars and Ortwin
Gerhard for helpful comments on the text and especially for pointing out the Sevenster
\& Kalnajs (2001) paper.

      Figure 3 is based on observations made with the {\it  Spitzer Space Telescope},
which is operated by JPL under contract~with NASA.
This work would not have been practical without extensive use of the 
NASA/IPAC Extragalactic Database (NED), which is operated by the Jet Propulsion 
Laboratory and the California Institute of Technology under contract with NASA.  
We also used the HyperLeda electronic database (Paturel et al.~2003) at 
{\tt http://leda.univ-lyon1.fr} and the image display tool SAOImage DS9 
developed by Smithsonian Astrophysical Observatory.  
% Figure 1 was adapted from the WIKISKY image database at {\tt http://www.wikisky.org}.
Finally, we made extensive use of NASA's Astrophysics Data System bibliographic 
services.

\end{document}